# Unidirectional Error Correcting Codes for Memory Systems: A Comparative Study

**Muzhir AL-ANI**[1] **and Qeethara AL-SHAYEA**[2]

[1] **Faculty of IT, Amman Arab University**
**Amman, Jordan**

[2] **MIS Department, Al-Zaytoonah University**
**Amman, Jordan**

**Abstract**
In order to achieve fault tolerance, highly reliable system often require the ability to detect errors as soon as they occur and prevent the speared of erroneous information throughout the system. Thus, the need for codes capable of detecting and correcting byte errors are extremely important since many memory systems use b-bit-per-chip organization. Redundancy on the chip must be put to make fault-tolerant design available. This paper examined several methods of computer memory systems, and then a proposed technique is designed to choose a suitable method depending on the organization of memory systems. The constructed codes require a minimum number of check bits with respect to codes used previously, then it is optimized to fit the organization of memory systems according to the requirements for data and byte lengths.
***Keywords:*** *Unidirctional Error Coding, Correcting Codes Design, Error Detection and Correcting and Error Constructing Codes.*

## 1. Introduction

In recent years, there has an increasing demand for efficient and reliable data transmission and storage systems. Fujiwara [1] insists that before designing a dependable system, we need to have enough knowledge of the system's faults, errors, and failures of the dependable techniques including coding techniques, and of the design process for practical codes.
Saitoh and Imai [2] represent codes that are capable of correcting byte and detecting multiple unidirectional bytes, but it is efficient code when b≤8. They also propose in [3] a code, but it is not efficient code for b≤8.
Zhang and Tu [4] propose a systematic t-EC/AUED codes which it's encoding and decoding is relatively easy, but it is efficient in the cases of t=1 and 2 and when k≤31.

S. Al-Bassam [5] presents an improved method to construct t error-correcting and all unidirectional error detecting codes (t-EC/AUED).
Umanesan and Fujiwara [6] propose a class of codes called Single t/b-error Correcting—Single b-bit byte Error Detecting codes which have the capability of correcting random t-bit errors occurring within a single b-bit byte and simultaneously indicating single b-bit byte errors.
Bose, Elmougy and Tallin [7] design some new classes of t-unidirectional error-detecting codes over $Z_m$.
Krishnan, Panigrahy and Parthasarathy [8] develop the error-correcting codes necessary to implement error-resilient ternary content addressable memories. They prove that the rate (ratio of data bits to total number of bits in the codewords) of the specialized error-correcting codes necessary for ternary content addressable memories cannot exceed 1/t, where t is the number of bit errors the code can correct.
Naydenova and Kløve [9] study codes that can correct up to t symmetric errors and detect all unidirectional errors. Biiinck and van Tilborg gave a bound on the length of binary such codes. They gave a generalization of this bound to arbitrary alphabet size. This generalized Biiinck-van Tilborg bound, combined with constructions, is used to determine some optimal binary and ternary codes for correcting t symmetric errors and detecting all unidirectional errors.
In computer memory, when data are stored in a byte-per-chip, byte errors may be occurring. When both one to zero and zero to one error may occur, but they do not occur simultaneously in a single byte, the errors are called a unidirectional byte error, which is a kind of byte error [10].

## 2. Coding Theory

The theory and practice of error-correction coding is concerned with protection of digital information against the errors that occur during data transmission or storage. Many ingenious error correcting techniques based on a vigorous mathematical theory have been developed and have many important and frequent applications. The current problem with any high-speed data communication system such as storage medium is how to control the errors that occur during storing data in storage medium. In order to achieve reliable communication, designers should develop good codes and efficient decoding algorithms [11].

There are three types of faults transient, intermittent, and permanent faults. Transient faults are likely to cause a limited number of symmetric errors or multiple unidirectional errors. Also, intermittent faults, because of short duration, are expected to cause a limited number of errors. On the other hand, permanent faults cause either symmetric or unidirectional errors, depending on the nature of the faults. The most likely faults in some of the recently developed LSI/VLSI, ROM, and RAM memories (such as the faults that affect address decoders, word lines, power supply, and stuck-fault in a serial bus, etc.) cause unidirectional errors. The number of unidirectional errors cause by the above mentioned faults can be fairly large [12].

The errors that can occur because of the noise are many and varied. However, they can be classified into three main types: symmetric, asymmetric, and unidirectional errors [7].

2.1 Error Control for Computer Main Memories

Error correcting codes have been used to enhance the reliability and data integrity of computer memory systems. The error correction can be incorporated in to the hardware.

In particular the class of single error-correcting and double error-detecting (SEC-DED) binary codes has been successfully used to correct and detect errors associated with failures in semiconductor memories. The most effective organization is the so-called 1 bit per chip organization. In this organization, all bits of a code word are stored in different chips. Any type of failures in a chip can corrupt at the most 1 bit of the code word. As long as the errors do not line up in the same code word, multiple errors in the memory are correctable. Large scale integration (LSI) and very large scale integration (VLSI) memory systems offer significant advantages in size, speed, and weight over earlier memory systems. These memories are normally packaged with multiple bit (or byte) per chip organization [13].

Coding techniques play a major role in segment the information in to m blocks each block of k-bit or it may be taken as a single block of length k (k=256, 512, 1024, 2048, 8192, 16384, 32768, 65536, 131072, 262144, 524288) according to the organized memory system in our research. BCH and RS code are two powerful approaches to error control coding in memory systems. The information segmented is the first step when information in a computer memory is written. Then this k-bit encoded in to n-bit called code word which consist of k-bit and r-bit parity check (n=k+r). This code word stored in memory.

The decoding method used to obtain the information k with no errors according to the coding technique when a code word fetched from the storage.

2.2 Reed-Solomon Codes (RS Codes)

A RS code is a class of non binary BCH codes. It is also a cyclic symbol error-correcting code. The RS code represent a very important class of algebraic error-correcting codes, which has been used for improving the reliability of compact disc, digital audio tape and other data storage systems [14]. Secure communications systems commonly use RS code as one method for protection against jamming. RS codes are also used for error control in the data storage systems, such as magnetic drums and photo digital storage systems.

A RS code is block sequence of finite field GF $(2^m)$ of $2^m$ binary symbols, where m is the number of bits per symbol. This sequence of symbols can be viewed as the coefficients of code polynomial $C(x)=c_0+c_1x+c_2x^2+\ldots+c_{n-1}x^{n-1}$ where the field elements $C_i$ are from $GF(2^m)$ [10].

A t-error-correcting RS code with symbols from $FG(2^m)$ has the following parameters:

Code length : $n=2^m-1$
Number of information : $k=n-2t$
Number of parity-check digits : $n-k=2t$
Minimum distance : $d_{min}=2t+1$

In the following, we shall consider Reed-Solomon codes with code symbols from the Galois field $GF(2^m)$. The generator polynomial of a t-error-correcting Reed-Solomon code of length $2^m-1$ is $g(x)=(x+\alpha)(x+\alpha^2)\ldots(x+\alpha^{2t})$, where $\alpha$ is a primitive element of $GF(2^m)$, and the coefficients $g_i$, $0 \leq l \leq 2t$ are also from $GF(2^m)$. An (n,k) RS code generated by g(x) is an (n,n-2t) cyclic code whose code vectors are multiples of g(x) [14,15].

Consider RS codes with symbols from $GF(2^m)$, where m is the number of bits per symbol.

56

Let $d(x) = c_{n-k}x^{n-k} + c_{n-k+1}x^{n-k+1} + \ldots + c_{n-1}x^{n-1}$ be the information polynomial and $p(x) = c_0 + c_1x + \ldots + c_{n-k-1}x^{n-k-1}$ be the check polynomial. Then the encoded RS code polynomial is expressed by:

$$c(x) = p(x) + d(x) \quad (1)$$

where $c_i, 0 \leq l \leq n-1$, are field elements in $GF(2^m)$. Thus, a vector of n symbols, $(c_0, c_1, \ldots, c_{n-1})$ is a code word if and only if its corresponding polynomial $c(x)$ is a multiple of the generator polynomial $g(x)$. The common method of encoding a cyclic code is to find $p(x)$ from $d(x)$ and $g(x)$, which results in an irrelevant quotient $q(x)$ and an important remainder $y(x)$. That is,

$$d(x) = q(x)g(x) + y(x) \quad (2)$$

Substituting Eq. (1) in to (2) gives:

$$c(x) = p(x) + q(x)g(x) + y(x) \quad (3)$$

If we define the check digits as the negatives of the coefficients of $y(x)$, i.e, $p(x) = -y(x)$, it follows that:

$$c(x) = q(x)g(x) \quad (4)$$

This ensures that the code polynomial $c(x)$ is multiple of $g(x)$. Thus, the RS encoder will perform the above division process to obtain the check polynomial $p(x)$ [14].

Theorem 1: A Reed-Solomon code is a maximum distance code, and the minimum distance is n-k+1.

This tells us that for fixed (n,k), no code can have a larger minimum distance than a RS code. This is often a strong justification for using RS codes. RS codes always have relatively short block length as compared to other cyclic codes over the same alphabet [16].

In decoding a RS code (or any non binary BCH code), the same three steps used for decoding a binary BCH code are required, in addition a fourth step involving calculation of the error value is required. The error value at the location corresponding to $B_1$ is given by the following equation:

$$e_{i1} = \frac{Z(\beta L^{-1})}{\pi^v_{I \approx}} \quad (5)$$

Where $z(x) = 1 + (s_1 + \sigma_1)x + (s_2 + \sigma_1 s_1 + \sigma_2)x^2 + \ldots + (s_v + \sigma s_{v-1} + \sigma_2 s_{v-2} + \ldots + \sigma_v)x^v$

The decoding method of RS code is worth mentioning because of its considerable theoretical interest, even though it is impractical [15].

## 3. Byte-Per-Chip Memory Organization

In many computer memory and VLSI circuits unidirectional errors are known to be predominant protection must be against combinations of unidirectional and random errors because random byte errors also appear from intermittent faults in memories. Thus it is very important to have such codes for protection of byte organized memories. Table (1) shows the parameters of modified RS code after shortening.

This code is optimal, thus it is the only SbEC-DbEC code with three check bytes but for a given size b(b<16) there is only one or two value of information.

Table 1: The parameters of shortened modified RS code

| b | n | n | k |
|---|---|---|---|
| 5 | 16 | 79 | 64 |
| 6 | 46 | 274 | 256 |
| 7 | 77 | 533 | 512 |
| 8 | 131 | 1048 | 1024 |
| 9 | 231 | 2075 | 2048 |
| 10 | 823 | 8222 | 8192 |
| 11 | 1493 | 16417 | 16384 |
| 12 | 2734 | 32804 | 32768 |
| 13 | 5045 | 65575 | 65536 |
| 14 | 9366 | 131114 | 131072 |
| 15 | 17480 | 262189 | 262144 |

Let the two codes whose $H_0$ matrices are denoted as Hv and Hw have minimum Hamming distance $d_{min}=4$ $GF(2^b)$, let $v_i$, i=0,1,…,n-1, denote a column vector in the matrix Hv. Preserving minimum distance, matrix Hw converted to matrix Hw having an all 'I' row vector. Next, this all 'I' row vector is removed from the matrix Hw, whose resultant matrix is now called Hw. Let $v_j$, j=0,1,…,m-1, denote a column vector of matrix Hw. The new code has a parity check matrix $H_1$ of the form that each column in it is defined by the following equation:

$$(C_{ij})^T = (V_i W_j) \quad (6)$$

i=0,1,…,n-1, and j=0,1,…,m-1. The $d_{min}$ of this code is four over $GF(2^b)$.

For example, let b=2, and Hw equal to

$$Hw = \begin{bmatrix} I & I & I & I & 0 & 0 \\ I & T & T^2 & 0 & I & 0 \\ I & T^2 & T & 0 & 0 & I \end{bmatrix} \quad (7)$$

where

$$I = \begin{bmatrix} 1 & 0 \\ 0 & 1 \end{bmatrix} \quad T^1 = \begin{bmatrix} 0 & 1 \\ 1 & 1 \end{bmatrix} \quad T^2 = \begin{bmatrix} 1 & 1 \\ 1 & 0 \end{bmatrix} \quad 0 = \begin{bmatrix} 0 & 0 \\ 0 & 0 \end{bmatrix}$$

This matrix can be converted to the new form that has top row vector which has all 'I' elements. This conversion can be carried out in the following manner.

The second row of Hw is multiplied by an arbitrary non zero element T^a. The multiplied result and the third row vector are added to the first row vector in Hw. If the added row vector has non zero element, each column can be normalized so that the first row element has a 'I' element. It can be derived that the number of T^a elements is $2^{b-1}$. If $T^a = T'$ is chosen then:

$$Hw' = \begin{bmatrix} I & I & I & I & 0 & 0 \\ I & T^2 & T^2 & 0 & I & 0 \\ T^2 & I & T^2 & 0 & 0 & I \end{bmatrix} \quad (8)$$

$$Hw'' = \begin{bmatrix} I & T^2 & T^2 & 0 & I & 0 \\ T^2 & I & T^2 & 0 & 0 & I \end{bmatrix} \quad (9)$$

Here $H_v$ as the $H_0$ matrix shown in Eq. (7) is adapted. An S2EC-D2ED code, whose $H_1$ matrix has five rows, can be constructed from $H_v$ and $H_{w''}$.

In the same manner, the SbEC-DbED codes whose $H_0$ matrices have odd number are obtained in the same way. If even number of rows is required (in this example), matrix Hw' can be shown as follows:

$$Hw' = \begin{bmatrix} I & I \\ 0 & I \end{bmatrix} \quad (10)$$

The code length N for the proposed codes is given as follows:

$$N = (2^b + 2)^{(r-1)/2} \quad \text{r:odd } (\geq 3) \quad (11)$$

$$N = 2(2^b + 2)^{(r-2)/2} \quad \text{r:even } (\geq 4) \quad (12)$$

Table 2: The parameters of SbEC-DbED RS codes

| b=5 | | b=6 | | b=7 | |
|---|---|---|---|---|---|
| r | n | k | n | k | n | k |
| 3 | 170 | 155 | 396 | 378 | 910 | 889 |
| 4 | 340 | 320 | 792 | 768 | 1820 | 1792 |
| 5 | 5780 | 5755 | 26136 | 26106 | 11300 | 118265 |
| 6 | 11560 | 11530 | 52272 | 52236 | 236600 | 236558 |

| b=8 | | b=9 | | b=10 | |
|---|---|---|---|---|---|
| r | n | k | n | k | n | k |
| 3 | 2064 | 2040 | 4626 | 4599 | 10260 | 10230 |
| 4 | 4128 | 4096 | 9252 | 9216 | 20520 | 20480 |

| b=11 | | b=12 | | b=13 | |
|---|---|---|---|---|---|
| r | n | k | n | k | n | k |
| 3 | 22550 | 22517 | 49176 | 49140 | 106522 | 106483 |
| 4 | 45100 | 45056 | 98352 | 98304 | 213044 | 212992 |

| b=14 | | b=15 | |
|---|---|---|---|
| r | n | k | n | k |
| 3 | 229404 | 229362 | 491550 | 491505 |
| 4 | 458808 | 458752 | | |

It is important to know that r is a parity check digits in bits, n is the code word length and k is the information length in all tables observer in this paper.
The Parameters of SbEC-DbED RS codes are illustrated in table (2). When the code is shortening table (3) is obtained.
It is obvious from comparing the parameters in table (1) with the parameters in table (3) that the parameters in table (3) are more efficient than the parameters in table (1).

Theorem 2 [17]: Let H be the parity check matrix of a (n,n-r) linear SbEC-DbED code over $GF(2_b)$. The (2n, 2n-r-1) linear code over $GF(2^b)$ defined by the parity check matrix H'.

$$H' = \begin{bmatrix} 0 & 0... & 0 & 1 & 1... & 1 \\ H & & & H & & \end{bmatrix} \quad (13)$$

Eq. (13) is a SbEC-DbED code.
Table (4) is obtained after applying theorem 2 to the parameters in table (2). Table (5) shows the results when the parameters are shorten.
After obvious comparison between the parameters in table (1) and parameters in table (3), we observe that there is no table with the best parameters for all value of k, so for the best parameters obtained table (6) is presented.
Since the two chip failure no longer takes place at the same time, these parameters can be used. Codes for only SbEC-DbED are proposed. So these codes can not recognize all the unidirectional errors which occur in b-bit-per-chip memory organization. Wherefore code that fits memory organized in b-bit-per-chip fashion, and 4<b<16 is constructed.

## 4. Conclusions

The most likely faults in many computer memories cause unidirectional errors, thus a detection of unidirectional errors is required. In addition, byte-error-correcting/detecting codes are useful for protection against byte errors which tend to occur when data are stored in byte-per-chip memory organization. A proposed technique for constructing SbEC-DbED codes is presented in this paper that can be practically applied to large capacity memory units. The obtained results indicate that the proposed technique is suitable and efficient for memory system to recognize unidirectional errors that occur in bit-per chip memory organization.

**Muzhir. Shaban Al-Ani** has received Ph. D. in Computer & Communication Engineering Technology, ETSII, Valladolid University, Spain, 1994. Assistant of Dean at Al-Anbar Technical Institute (1985). Head of Electrical Department at Al-Anbar Technical Institute, Iraq (1985-1988), Head of Computer and Software Engineering Department at Al-Mustansyria University, Iraq (1997-2001), Dean of Computer Science (CS) & Information System (IS) faculty at University of Technology, Iraq (2001-2003). He joined in 15 September 2003 Electrical and Computer Engineering Department, College of Engineering, Applied Science University, Amman, Jordan, as Associated Professor. He joined in 15 September 2005 Management Information System Department, Amman Arab University, Amman, Jordan, as Associated Professor, then he joined computer science department in 15 September 2008 at the same university.

**Qeethara Kadhim Abdul Rahman Al-Shayea** has received Ph. D. in Computer Science, Computer Science Department, University of Technology, Iraq, 2005.   She received her M.Sc degree in Computer Science, Computer Science Department from University of Technology, Iraq, 2000. She has received her High Diploma degree in information Security from Computer Science Department, University of Technology, Iraq, 1997. She joined in 15 September (2001-2006), Computer Science Department, University of Technology, Iraq as assistant professor. She joined in 15 September 2006, Department of Management Information Systems Faculty of Economics & Administrative Sciences Al-Zaytoonah University of Jordan as assistant professor. She is interested in Coding Theory, Computer Vision and Artificial Intelligence.

Table 3: The parameters of shortened SbEC-DbED RS codes

| b=5 | | | b=6 | | | b=7 | | |
|---|---|---|---|---|---|---|---|---|
| r | n | k | r | n | k | r | n | k |
| 3 | 47 | 32 | 3 | 82 | 64 | 3 | 149 | 128 |
|   | 79 | 64 |   | 146 | 128 |   | 277 | 256 |
|   | 143 | 128 |   | 274 | 256 |   | 533 | 512 |
| 4 | 276 | 256 | 4 | 536 | 512 | 4 | 1052 | 1024 |
| 5 | 537 | 512 | 5 | 1054 | 1024 | 5 | 2083 | 2048 |
|   | 1049 | 1024 |   | 2078 | 2048 |   | 4131 | 4096 |
|   | 2073 | 2048 |   | 4126 | 4096 |   | 8227 | 8192 |
|   | 4121 | 4096 |   | 8222 | 8192 |   | 16419 | 16384 |
| 6 | 8222 | 8192 |   | 16414 | 16384 |   | 32803 | 32768 |
| 7 | 16419 | 16384 | 6 | 32804 | 32768 |   | 65571 | 65536 |
|   | 32803 | 32768 | 7 | 65578 | 65536 | 6 | 131114 | 131072 |
|   | 65571 | 65536 |   | 131114 | 131072 | 7 | 262193 | 262144 |
|   | 131107 | 131072 |   | 262186 | 262144 |   | 524337 | 524288 |
| 8 | 262184 | 262144 |   | 524330 | 524288 |   |   |   |
| 9 | 524333 | 524288 |   |   |   |   |   |   |
| b=8 | | | b=9 | | | b=10 | | |
| r | n | k | r | n | k | r | n | k |
| 3 | 280 | 256 | 3 | 539 | 512 | 3 | 1054 | 1024 |
|   | 536 | 512 |   | 1051 | 1024 |   | 2078 | 2048 |
| | 1048 | 1024 | | 2075 | 2048 | | 4126 | 4096 |
| 4 | 2080 | 2048 | | 4123 | 4096 | | 8222 | 8192 |
| | 4128 | 4096 | 4 | 8228 | 8192 | 4 | 16424 | 16384 |
| 5 | 8232 | 8192 | 5 | 16429 | 16384 | 5 | 32818 | 32768 |
| | 16424 | 16384 | | 32813 | 32768 | | 65586 | 65536 |
| | 32808 | 32768 | | 65581 | 65536 | | 131122 | 131072 |
| | 65576 | 65536 | | 131117 | 131072 | | 262194 | 262144 |
| | 131112 | 131072 | | 262189 | 262144 | | 524338 | 524288 |
| | 262184 | 262144 | | 524333 | 524288 | | | |
| | 524328 | 524288 | | | | | | |
| | **b=11** | | | **b=12** | | | **b=13** | |
| r | n | k | r | N | k | r | n | K |
| 3 | 2081 | 2048 | 3 | 4132 | 4096 | 3 | 8231 | 8192 |
| | 4129 | 4096 | | 8228 | 8192 | | 16423 | 16384 |
| | 8225 | 8192 | | 16420 | 16384 | | 32807 | 32768 |
| | 16417 | 16384 | | 32804 | 32768 | | 65575 | 65536 |
| 4 | 32812 | 32768 | 4 | 65584 | 65536 | 4 | 131124 | 131072 |
| 5 | 65591 | 65536 | | 131132 | 131072 | 5 | 262209 | 262144 |
| | 131127 | 131072 | | 262204 | 262144 | | 524353 | 524288 |
| | 262199 | 262144 | | 524348 | 524288 | | | |
| | 524343 | 524288 | | | | | | |
| | **b=14** | | | **b=15** | | | | |
| r | n | k | r | N | k | | | |
| 3 | 16426 | 16384 | 3 | 32813 | 32768 | | | |
| | 32810 | 32768 | | 65581 | 65536 | | | |
| | 65578 | 65536 | | 131117 | 131072 | | | |
| | 131114 | 131072 | | 262189 | 262144 | | | |
| 4 | 262200 | 262144 | 4 | 524348 | 524288 | | | |
| | 524344 | 524288 | | | | | | |

Table 4: The parameters of new SbEC-DbED RS codes

| **b=5** | | | **b=6** | | | **b=7** | | |
|---|---|---|---|---|---|---|---|---|
| r | n | k | r | n | K | r | n | k |
| 16 | 340 | 324 | 19 | 792 | 773 | 22 | 1820 | 1798 |
| 21 | 680 | 659 | 25 | 1584 | 1559 | 29 | 3640 | 3611 |
| 26 | 11560 | 11534 | 31 | 52272 | 52241 | 36 | 236600 | 236564 |
| **b=8** | | | **b=9** | | | **b=10** | | |
| r | n | k | r | n | K | r | n | k |
| 25 | 4128 | 4103 | 28 | 9252 | 9224 | 31 | 20520 | 20489 |
| 33 | 8256 | 8223 | 37 | 18504 | 18467 | 41 | 41040 | 40999 |
| **b=11** | | | **b=12** | | | **b=13** | | |
| r | n | k | r | n | K | r | n | k |
| 34 | 45100 | 45066 | 37 | 98352 | 98315 | 40 | 213044 | 213004 |
| 45 | 90200 | 90155 | 49 | 196704 | 196655 | 53 | 426088 | 426035 |
| **b=14** | | | **b=15** | | | | | |
| r | n | k | r | n | K | | | |
| 43 | 458808 | 458765 | 46 | 983100 | 983054 | | | |

Table 5: The parameters of shortened new SbEC-DbED RS codes

| **b=5** | | | **b=6** | | | **b=7** | | |
|---|---|---|---|---|---|---|---|---|
| r | n | k | r | n | k | r | n | k |
| 16 | 48 | 32 | 19 | 83 | 64 | 22 | 150 | 128 |
| | 80 | 64 | | 147 | 128 | | 278 | 256 |
| | 144 | 128 | | 275 | 256 | | 534 | 512 |

|    | 272    | 256    |    | 531    | 512    |    | 1046   | 1024   |
|----|--------|--------|----|--------|--------|----|--------|--------|
| 21 | 533    | 512    | 25 | 1049   | 1024   | 29 | 2077   | 2048   |
| 26 | 1050   | 1024   | 31 | 2079   | 2048   | 36 | 4132   | 4096   |
|    | 2074   | 2048   |    | 4127   | 4096   |    | 8228   | 8192   |
|    | 4122   | 4096   |    | 8223   | 8192   |    | 16420  | 16384  |
|    | 8218   | 8192   |    | 16415  | 16384  |    | 32804  | 32768  |
| 31 | 16415  | 16384  |    | 326799 | 32768  |    | 65572  | 65536  |
| 36 | 32804  | 32768  | 37 | 65573  | 65536  |    | 131104 | 131072 |
|    | 65572  | 65536  | 43 | 131115 | 131072 | 43 | 262187 | 262144 |
|    | 131108 | 131072 |    | 262187 | 262144 | 50 | 524338 | 524288 |
|    | 262180 | 262144 |    | 524331 | 524288 |    |        |        |
|    | 524329 | 524288 |    |        |        |    |        |        |

|    | b=8    |        |    | b=9    |        |    | b=10   |        |
|----|--------|--------|----|--------|--------|----|--------|--------|
| r  | n      | k      | r  | N      | K      | r  | n      | k      |
| 25 | 281    | 256    | 28 | 540    | 512    | 31 | 1055   | 1024   |
|    | 537    | 512    |    | 1052   | 1024   |    | 2079   | 2048   |
|    | 1049   | 1024   |    | 2076   | 2048   |    | 4127   | 4096   |
|    | 2073   | 2048   |    | 4124   | 4096   |    | 8223   | 8192   |
|    | 4121   | 4096   |    | 8220   | 8192   |    | 16415  | 16384  |
| 33 | 8225   | 8192   | 37 | 16421  | 16384  | 41 | 32809  | 32768  |
| 41 | 16425  | 16384  | 46 | 32814  | 32768  | 51 | 65587  | 65536  |
|    | 32809  | 32768  |    | 65582  | 65536  |    | 131123 | 131072 |
|    | 65577  | 65536  |    | 131118 | 131072 |    | 262195 | 262144 |
|    | 131113 | 131072 |    | 262190 | 262144 |    | 524339 | 524288 |
|    | 262185 | 262144 |    | 524334 | 524288 |    |        |        |
|    | 524329 | 524288 |    |        |        |    |        |        |

|    | b=11   |        |    | b=12   |        |    | b=13   |        |
|----|--------|--------|----|--------|--------|----|--------|--------|
| r  | n      | k      | r  | n      | k      | r  | n      | k      |
| 34 | 2080   | 2048   | 37 | 4133   | 4096   | 40 | 8232   | 8192   |
|    | 4130   | 4096   |    | 8229   | 8192   |    | 16424  | 16384  |
|    | 8226   | 8192   |    | 16421  | 16384  |    | 32808  | 32768  |
|    | 16418  | 16384  |    | 32805  | 32768  |    | 65576  | 65536  |
|    | 32802  | 32768  |    | 65573  | 65536  |    | 131112 | 31072  |
| 45 | 65581  | 65536  | 49 | 131121 | 131072 | 53 | 262197 | 262144 |
| 56 | 131128 | 131072 | 61 | 262205 | 262144 | 66 | 524354 | 524288 |
|    | 262200 | 262144 |    | 524349 | 524288 |    |        |        |
|    | 524344 | 524288 |    |        |        |    |        |        |

|    | b=14   |        |    | b=15   |        |
|----|--------|--------|----|--------|--------|
| r  | n      | k      | r  | n      | k      |
| 43 | 16427  | 16384  | 46 | 32814  | 32768  |
|    | 32811  | 32768  |    | 65582  | 65536  |
|    | 65579  | 65536  |    | 131118 | 131072 |
|    | 131115 | 131072 |    | 262190 | 262144 |
|    | 262187 | 262144 |    | 524334 | 524288 |
|    | 524345 | 524288 |    |        |        |

Table 6: Best parameters obtained

| b=5        |       | b=6        |           | b=7        |           |
|------------|-------|------------|-----------|------------|-----------|
| (n,k)      | Table | (n,k)      | Table no. | (n,k)      | Table no. |
| (47,32)    | (3)   | (82,64)    | (3)       | (149,128)  | (3)       |
| (79,64)    | (1,3) | (146,128)  | (3)       | (277,256)  | (3)       |
| (143,128)  | (1,3) | (274,256)  | (1,3)     | (533,512)  | (1,3)     |
| (272,256)  | (5)   | (531,512)  | (1)       | (1046,1024)| (5)       |
| (533,512)  | (5)   | (1049,1024)| (1)       | (2077,2048)| (5)       |
| (1049,1024)| (3)   | (2078,2048)| (3)       | (4131,4096)| (3)       |
| (2073,2048)| (3)   | (4126,4096)| (3)       | (8227,8192)| (3)       |

| (n,k) | Table no. | (n,k) | Table no. | (n,k) | Table no. |
|---|---|---|---|---|---|
| (4121,4096) | (3) | (8222,8192) | (3) | (10419,16384) | (3) |
| (8218,8192) | (5) | (16414,16384) | (3) | (32803,32768) | (3) |
| (16415,16384) | (5) | (32799,32768) | (5) | (65571,65536) | (3) |
| (32803,32768) | (3) | (65573,65536) | (5) | (131108,131072) | (5) |
| (65571,65536) | (3) | (131114,131072) | (3) | (262187,262144) | (5) |
| (131107,131072) | (3) | (262186,262144) | (3) | (524377,524288) | (3) |
| (262180,262144) | (5) | (524330,524288) | (3) | | |
| (524329,524288) | (5) | | | | |
| **b=8** | | **b=9** | | **b=10** | |
| (n,k) | Table no. | (n,k) | Table no. | (n,k) | Table no. |
| (280,256) | (3) | (539,512) | (3) | (1054,1024) | (3) |
| (536,512) | (3) | (1051,1024) | (3) | (2078,2048) | (3) |
| (1048,1024) | (1,3) | (2075,2048) | (1,3) | (4126,4096) | (3) |
| (2073,2048) | (5) | (4123,4096) | (1,3) | (8222,8192) | (1,3) |
| (4121,4096) | (5) | (8220,8192) | (5) | (16415,16384) | (5) |
| (8225,8192) | (5) | (16421,16384) | (5) | (32809,32768) | (5) |
| (16424,16384) | (3) | (32813,32768) | (3) | (65586,65536) | (3) |
| (32808,32768) | (3) | (65581,65536) | (3) | (131122,131072) | (3) |
| (65576,65536) | (3) | (131117,131072) | (3) | (262194,262144) | (3) |
| (131112,131072) | (3) | (262189,262144) | (3) | (524338,524288) | (3) |
| (262184,262144) | (3) | (524333,524288) | (3) | | |
| (524328,524288) | (3) | | | | |
| **b=11** | | **b=12** | | **b=13** | |
| (n,k) | Table no. | (n,k) | Table no. | (n,k) | Table no. |
| (2081,2048) | (3) | (4132,4096) | (3) | (8231,8192) | (3) |
| (4129,4096) | (3) | (8228,8192) | (3) | (16423,16384) | (3) |
| (8225,8192) | (3) | (16420,16384) | (3) | (32807,32768) | (3) |
| (16417,16384) | (1,3) | (32804,32768) | (1,3) | (65575,65536) | (1,3) |
| (32802,32768) | (5) | (65573,65536) | (5) | (131112,131072) | (5) |
| (65581,65536) | (5) | (131121,131072) | (5) | (262197,262144) | (5) |
| (131127,131072) | (3) | (262204,262144) | (3) | (524353,524288) | (3) |
| (262199,262144) | (3) | (524348,524288) | (3) | | |
| (524343,524288) | (3) | | | | |
| **b=14** | | **b=15** | | | |
| (n,k) | Table no. | (n,k) | Table no. | | |
| (16426,16384) | (3) | (32813,32768) | (3) | | |
| (32810,32768) | (3) | (65581,65536) | (3) | | |
| (65578,65536) | (3) | (131117,131072) | (3) | | |
| (131114,131072) | (1,3) | (262189,262144) | (3) | | |
| (262187,262144) | (5) | (524334,524288) | (5) | | |
| (524344,524288) | (3) | | | | |